\documentclass[prl,twocolumn,floatfix,epsf,amssymb,longbibliography,superscriptaddress,aps,fleqn]{revtex4-1}
\usepackage{graphicx,subfigure,amsmath,colordvi,times}
\usepackage{braket}
\usepackage[usenames]{color}
\usepackage[english]{babel}
\usepackage{amstext}
\usepackage{amsfonts}
\usepackage{amssymb}
\usepackage{bm}
\usepackage{float}
\usepackage{xcolor} 
\usepackage{hyperref}
\hypersetup{pdfborder=0 0 0,colorlinks=true,citecolor=blue,linkcolor=blue}

\def\urlprefix{}
\def\url#1{}

\begin{document}

\title{Three-dimensional electron-hole superfluidity in a superlattice close to room temperature}

\author{M. Van der Donck}
\email{matthias.vanderdonck@uantwerpen.be}
\affiliation{Department of Physics, University of Antwerp, Groenenborgerlaan 171, 2020 Antwerp, Belgium}
\author{S. Conti}
\affiliation{Department of Physics, University of Antwerp, Groenenborgerlaan 171, 2020 Antwerp, Belgium}
\affiliation{Physics Division,  School of Science \& Technology, Universit\`a di Camerino, 62032 Camerino (MC), Italy}
\author{A. Perali}
\affiliation{Supernano Laboratory, School of Pharmacy, Universit\`a di Camerino, 62032 Camerino (MC), Italy}
\affiliation{ARC Centre of Excellence for Future Low Energy Electronics Technologies, 
School of Physics, The University of New South Wales, Sydney, N.S.W. 2052, Australia}
\author{A. R. Hamilton}
\affiliation{ARC Centre of Excellence for Future Low Energy Electronics Technologies, 
School of Physics, The University of New South Wales, Sydney, N.S.W. 2052, Australia}
\author{B. Partoens}
\affiliation{Department of Physics, University of Antwerp, Groenenborgerlaan 171, 2020 Antwerp, Belgium}
\author{F. M. Peeters}
\affiliation{Department of Physics, University of Antwerp, Groenenborgerlaan 171, 2020 Antwerp, Belgium}
\author{D. Neilson}
\affiliation{Department of Physics, University of Antwerp, Groenenborgerlaan 171, 2020 Antwerp, Belgium}
\affiliation{ARC Centre of Excellence for Future Low Energy Electronics Technologies, 
School of Physics, The University of New South Wales, Sydney, N.S.W. 2052, Australia}
 
\begin{abstract}
Although there is strong theoretical and experimental evidence for electron-hole superfluidity in separated
sheets of electrons and holes at low $T$, extending superfluidity to high $T$ is limited by strong 2D fluctuations and
Kosterlitz-Thouless effects. We show this limitation can be overcome using a superlattice of alternating electron- and
hole-doped semiconductor monolayers. The superfluid transition in a 3D superlattice is not
topological, and for strong electron-hole pair coupling, the transition temperature $T_c$ can be at room temperature.
As a quantitative illustration, we show $T_c$ can reach $270$ K for a superfluid in a realistic superlattice of transition metal dichalcogenide monolayers.
\end{abstract}
\maketitle



It was predicted half a century ago that bound pairs of electrons and holes (excitons) in a semiconductor 
should quantum condense at low temperatures\cite{Keldysh1965}. 
To prevent fast electron-hole (e-h) recombination, the electrons and holes can be confined 
in two spatially separated two-dimensional (2D) layers\cite{Lozovik1975}. 
At atomically small layer separations, the attractive Coulomb interaction is strong and  e-h  binding energies in excess of $1000$ K 
have been demonstrated\cite{Rivera2018}. 
Under appropriate conditions, these indirect excitons are predicted to form a superfluid condensate with a large energy gap\cite{Min2008,Perali2013}. 
Enhanced tunneling has been observed in e-h double-bilayers\cite{Burg2018} at transition temperatures $T_c\sim 1$ K. 
Such enhancement of tunneling is a strong indication of superfluidity or Bose-Einstein condensation (BEC)\cite{Efimkin2019}.
A dramatic increase in $T_c$ was recently reported with the observation of enhanced tunneling up to $T_c\sim 100$ K 
in a double-monolayer transition metal dichalcogenide (TMD) heterostructure\cite{Wang2019,Chaves2019}, 
in agreement with recent predictions\cite{Conti2019b}.   

One might reasonably expect that the transition temperature could be further increased up to the limit set 
by the large pair binding energies  $\sim 1000$ K and the large superfluid gaps $\gg 300$ K.  
However any further increase of the transition temperature in these quasi-2D systems is blocked by the  Mermin–Wagner theorem\cite{Mermin1966,Hohenberg1967}.
Thus the maximum transition temperature is not limited by the e-h binding energy or superfluid gap, 
but by a Berezinskii-Kosterlitz-Thouless (BKT) topological transition\cite{Kosterlitz1973}.  
The transition temperature $T^{BKT}$ is proportional to the carrier density, so it does not increase with coupling strength.  
Increasing $T^{BKT}$ by increasing the density is not possible, because strong screening of the e-h Coulomb interactions at high densities
kills the superfluidity\cite{Lozovik2012,Perali2013}.

Here, we overcome the restrictions associated with Mermin-Wagner and exploit the strong e-h coupling,  
by considering superfluidity in a three-dimensional (3D) superlattice, consisting of a stack   
of alternating electron and hole monolayers.  In a 3D system, strong e-h coupling and the associated large superfluid gaps 
can lead to superfluid transitions at room temperature.    
We focus specifically on a superlattice of alternating electron-doped and hole-doped monolayers 
of the transition metal dichalcogenides $n$-WS$_2$ and $p$-WSe$_2$, but the  
approach would work for other systems of stacked e-h 2D layers. 
We note there are already many examples of superlattice-based superconductors\cite{Croitoru2012}, including 
the high-$T_c$ cuprates\cite{Ioffe1999,Clarke1997}.

\begin{figure}
\begin{center}
\includegraphics[trim=2cm 0.5cm 0.5cm 0cm, clip=true, angle=0,width=1\columnwidth] {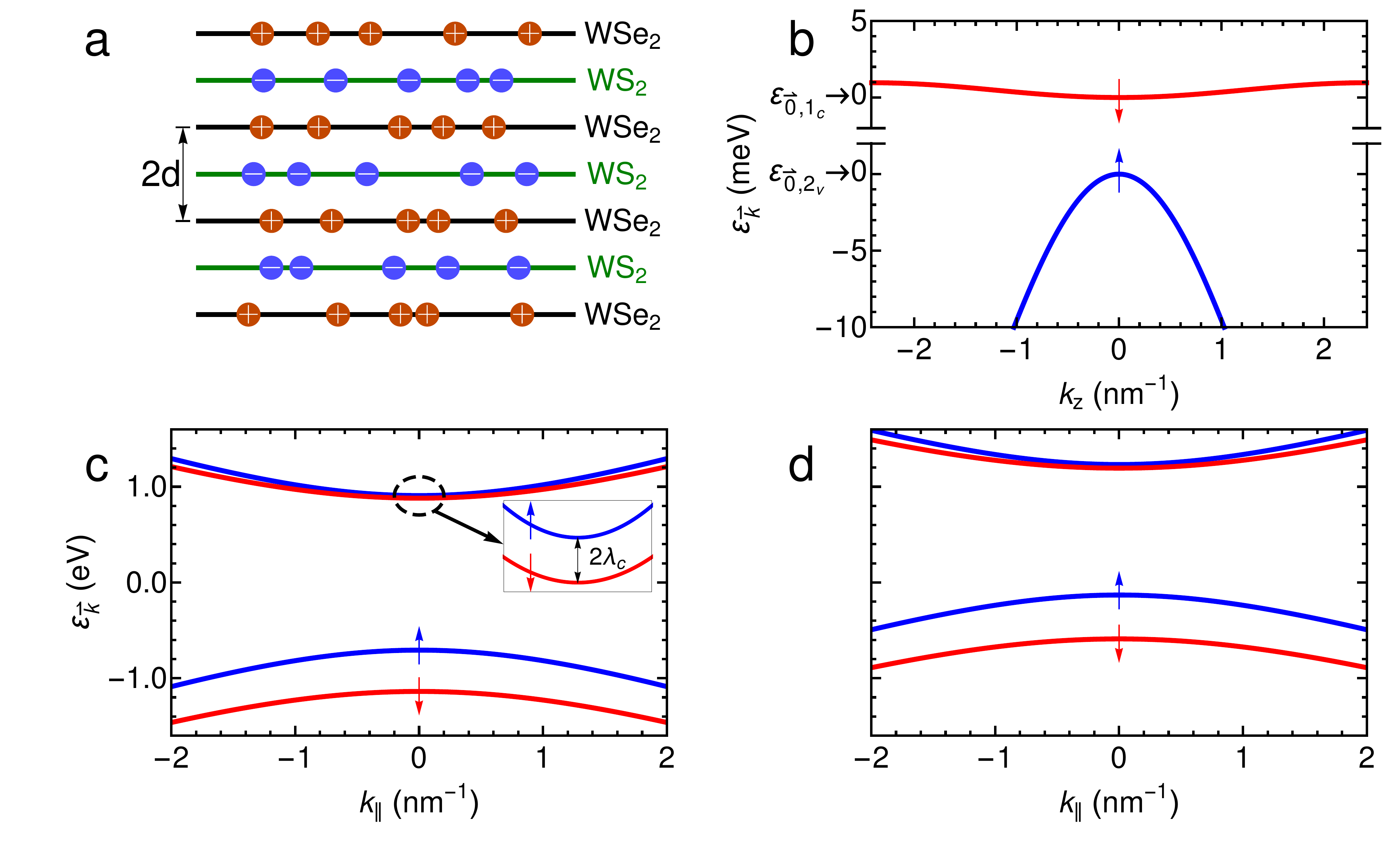}
\end{center}
\caption{
(Color online) 
(a) Schematic illustration of the WS${_2}$/WSe${_2}$ heterostructure superlattice of periodicity $2{d}$, of alternating monolayers of the two different TMDs, 
one type $n$-doped (green lines) and the other $p$-doped (black lines). 
(b) Lowest conduction band and highest valence band of the 
superlattice as a function of ${k_z}$, expressed relative to the centre of the $K$ valley.
Blue and red bands are spin up and spin down bands, respectively.  For the $K'$ valley, the spins are reversed.
(c) Bands predominantly associated with WS${_2}$ as a function of $\bm{k_\parallel}$.
Inset shows a close up of the two spin-split WS${_2}$ conduction bands,  
separated by ${2\lambda_c=27}$ meV. 
(d) Bands predominantly associated with WSe${_2}$ as a function of $\bm{k_\parallel}$.
}
\label{Bandstructure}
\end{figure}

Figure \ref{Bandstructure}(a) schematically shows the infinite superlattice of alternating $n$- and $p$-doped monolayers of two different TMDs, 
indicated by green and black lines.  
Within each monolayer, a layer of W transition metal atoms   
is sandwiched between two layers of S or Se chalcogen atoms.
We consider an AA stacked superlattice of WS$_2$ and  WSe$_2$ monolayers,
with the tungsten atoms horizontally aligned, and the chalcogen atoms horizontally aligned. 
For this stacking, the superlattice has a direct band gap\cite{Terrones2013}.
Electrons and holes  generated by the alternate $n$- and $p$-doping of the monolayers form bound pairs.
The WS$_2$/WSe$_2$ band alignment is type-II, which keeps the
electrons and the holes spatially separated in their monolayers.
This ensures  long lifetimes for the interlayer excitons:  
in a related double-monolayer MoSe$_2$/WSe$_2$ system, 
optically generated interlayer excitons with lifetimes $\sim 1.8$ ns 
have been observed\cite{Rivera2015}.


We start with a hybrid continuum-tight binding approach to determine the band structure of the superlattice.
The low-energy single-particle Hamiltonian for this superlattice, valid in the $K$ and $K'$ valleys, can be written as, 
\begin{equation}
\label{singleham}
H_{\vec{k},s,\tau} = \begin{pmatrix}
H_{s,\tau}^1(k_x,k_y) & {\cal{T}}(k_z) \\
T^{\dag}(k_z) & H_{s,\tau}^2(k_x,k_y)+\delta_bI_2
\end{pmatrix}\ .
\end{equation}
The indices are $s=\pm1$ for  spin, and $\tau=+1$ and $-1$ for valley $K$ and $K'$, respectively. 
We will represent 2D vectors in the $x$-$y$ space of the monolayer planes as $\bm{k}_\parallel$,
and vectors in 3D space as $\vec{k}\equiv (\mathbf{k}_\parallel,k_z)$, 
with the $z$-direction perpendicular to the monolayers. 
The $\bm{k}_\parallel$ momentum vectors are expressed relative to the center of the $K$ or $K'$ valley. 

The Hamiltonian for the type $\ell=1$ (WS$_2$) or type $\ell=2$ (WSe$_2$) monolayer can be expressed for low energies in a Bloch basis, 
one for each type TMD monolayer, comprising    
the transition metal atomic orbitals for the lowest conduction band, $d_0$, and the highest valence band, 
$d_{\pm2}$ (the plus and minus correspond to the $K$ and $K'$ valley, respectively)\cite{Xiao2012}.
\begin{equation}
\label{monoham}
H_{s,\tau}^\ell(k_x,k_y) = \begin{pmatrix}
\frac{E^g_\ell}{2}+\lambda_{c,\ell}s\tau & a_\ell t_\ell (\tau k_x-ik_y) \\
a_\ell t_\ell (\tau k_x+ik_y) & -\frac{E^g_\ell}{2}+\lambda_{v,\ell}s\tau
\end{pmatrix}\ .
\end{equation}
$a_\ell$ is the lattice constant of the type $\ell$ monolayer, 
$t_\ell$ the intralayer hopping parameter, and $E^g_\ell$ the band gap.  
$\lambda_{c,\ell}$ and $\lambda_{v,\ell}$ are the spin-orbit coupling strengths 
in the conduction and valence bands. 
Values of these parameters are found in Table S1 in the Supplementary Material\cite{Supplemental3DTMD}. 
A bias potential between the two different TMDs, $\delta_b=0.412$ eV, ensures that the band alignment agrees with Ref.\ \citenum{Kang2013}.  

In Eq.\ \eqref{singleham}, $ {\cal{T}}(k_z)$ is the interlayer part of the Hamiltonian, 
%
\begin{equation}
\label{interlayer}
 {\cal{T}}(k_z) = \begin{pmatrix}
2t_c\cos(k_zd) & 0 \\
0 & 2t_v\cos(k_zd)
\end{pmatrix}\ ,
\end{equation}
where $d=0.65$ nm is the distance between monolayers.  $t_c$ and $t_v$ are the interlayer hopping parameters 
between the conduction band $d_0$-states and the valence band $d_{\pm2}$-states
of the opposite monolayers. 

For AA stacking, the interlayer nearest neighbors have the same in-plane coordinates, 
so the interlayer hopping between the $d_0$-states does not vanish. 
The coupling strength between the $d_{\pm2}$-states is almost identical for AA and AB stacking\cite{Wang2017}.  
From bilayer MoS$_2$ we know that the coupling strength between the $d_0$-states is $\sim \tfrac{1}{7}$ of the coupling strength 
between the $d_{\pm2}$-states\cite{Wang2017}.   Since the coupling strength is  determined only by the type of orbitals 
and the spatial separation, which is the same for all TMDs, we will assume $t_c=\tfrac{1}{7}t_v$ as a general relation. 
For TMD heterostructures, the effective hopping parameter is assumed to be given by the average value 
of the hopping parameters of each of the two TMDs, in general a good approximation\cite{Wang2017}. 
For our  WS$_2$/WSe$_2$ superlattice, the transition metal atoms of the  TMD monolayers are the same, 
making this an even better approximation.

The energy spectrum shown in Fig.\ \ref{Bandstructure}(b)-(d) and the corresponding eigenstates are obtained by numerically solving 
the eigenvalue equation of the $4\times 4$ Hamiltonian, Eq.\ \eqref{singleham}. 
For  a given spin and valley quantum number, the single-particle eigenstate for energy band 
$\beta$ is $\ket{\psi_{\vec{k},\beta}}$.
For the WS$_2$ monolayer conduction band, we need consider only the lowest conduction band, 
with spin down (up) for the $K$ ($K'$) valley (see Fig.\ \ref{Bandstructure}c),  
since the band above will start to fill only for $T\gtrsim 300$ K. 
We label the corresponding superlattice band $\beta=1_c$, referring to the dominant component in Eq.\ (S1) 
in the Supplementary Material\cite{Supplemental3DTMD}.   
Similarly, for the valence band of the WSe$_2$ monolayer, the very large spin splitting 
means that we need consider only the highest valence band, with spin up (down) for the $K$ ($K'$) valley.  
We label the corresponding superlattice band $\beta=2_v$.  
Because of the spin polarization in the valleys, 
the number of flavors for the electrons and holes comes only from the valley degeneracy, $g_v=2$.
Figure\ \ref{Bandstructure}(b) shows the lowest conduction band and highest valence band of the WS${_2}$/WSe${_2}$ heterostructure superlattice 
as a function of the perpendicular wave vector component ${k_z}$, expressed relative to the center of the $K$ valley. 
Blue and red bands are spin up and spin down bands, respectively.  For the $K'$ valley, the spins are reversed.
Figure\ \ref{Bandstructure}(c) and (d) show the bands associated predominantly with WS${_2}$
and WSe${_2}$, respectively, 
as a function of the in-plane wave vector component $\bm{k_\parallel}$, again relative to the $K$ valley.


We will evaluate the bare Coulomb interaction matrix elements 
$\braket{\psi_{\vec{\kappa}',\alpha'}\psi_{\vec{k}',\beta'}|V|\psi_{\vec{\kappa},\alpha}\psi_{\vec{k},\beta}}$ 
for e-h scattering between the $\ket{\psi_{\vec{k},\beta}}$ eigenstates of the superlattice, 
with $V(r)=-{e^2}/\left({4\pi\varepsilon_r\varepsilon_0r}\right)$.
The dielectric constant $\varepsilon_r$ accounts for static screening effects of both ions and the filled valence bands. 
For bulk WS$_2$ $\varepsilon_r=\sqrt{\varepsilon_z\varepsilon_{\parallel}}=9.9$, 
and for WSe$_2$ $\varepsilon_r=11.2$\cite{Laturia2018}. 
In the limit of no hybdridization between the different TMD types, the system 
would effectively consist of two decoupled bulk TMDs with an interlayer distance 
twice that of their normal bulk forms. It is shown in Ref.\ \citenum{Berkelbach2013} 
that the dielectric constant of MoS$_2$ is approximately halved when the interlayer distance is doubled. 
For the WS$_2$/WSe$_2$ superlattice, we therefore take as the value of the dielectric constant 
for the  heterostructure superlattice $\varepsilon_r= 5.5$, half of the average of the two bulk TMDs.
While the Keldysh potential\cite{Keldysh1979} applies for monolayer TMDs, 
here the nature of the interactions in $\braket{\psi_{\vec{\kappa}',\alpha'}\psi_{\vec{k}',\beta'}|V|\psi_{\vec{\kappa},\alpha}\psi_{\vec{k},\beta}}$ 
is 3D and  the average interparticle distances for the densities we are considering are much 
larger than the small distance between layers. 

The interaction between electrons and holes from same type TMD monolayers is given by\cite{Fetter1974,Kotov2012} , 
\begin{equation}
\label{Vintra}
\!\!\!\!\!\!V^{(0)}(\bm{q}_{\parallel},q_z) \!=\! \frac{-e^2}{4\pi\varepsilon_r\varepsilon_0NA}\frac{2\pi}{q_{\parallel}}
\!\left[\frac{\sinh(2q_{\parallel}d)}{\cosh(2q_{\parallel}d)-\cos(2q_zd)}\right] \!\!\!
\end{equation}
(for details see discussion in the Supplementary Material\cite{Supplemental3DTMD}). 
Equation \eqref{Vintra} passes between the correct 2D and 3D limits (see Fig.\ S1 in the Supplementary Material\cite{Supplemental3DTMD}). 
In the limit $d\rightarrow \infty$, the rightmost term is equal to unity, 
and we recover the 2D interaction potential for $N$ layers of surface area $A$.
In the limit $d\rightarrow 0$, a Taylor expansion of the trigonometric functions 
transforms the rightmost term to $2q_{\parallel}/(2d(q_{\parallel}^2+q_z^2))$, 
thus recovering the 3D interaction potential for volume $(AN2d)$. 

For electrons and holes from different type TMD monolayers, we find that the interaction is,
\begin{equation}
\label{Vinter}
\!\!\!\!\!\!V^{(d)}(\bm{q}_{\parallel},q_z) \!=\! \frac{-e^2}{4\pi\varepsilon_r\varepsilon_0NA}\frac{2\pi}{q_{\parallel}}
\!\left[\frac{2\sinh(q_{\parallel}d)\cos(q_zd)}{\cosh(2q_{\parallel}d)-\cos(2q_zd)}\right] \!\!\!
\end{equation}
In the limit $d\rightarrow 0$, Eq.\ \eqref{Vinter} reduces to the standard 3D interaction potential, 
while the limit $d\rightarrow \infty$ introduces the familiar factor $2e^{-q_{\parallel}d}$.


When evaluating $\braket{\psi_{\vec{\kappa}',\alpha'}\psi_{\vec{k}',\beta'}|V|\psi_{\vec{\kappa},\alpha}\psi_{\vec{k},\beta}}$, 
it suffices to consider the dominant intraband interactions: $\alpha=\alpha'=1_c$ and $\beta=\beta'=2_v$ 
because of the large energy band gaps.   
For the superfluid calculations, e-h pairs with zero center of mass momentum are required  
for which the interaction is,
\begin{eqnarray}
\label{interintra}
\braket{\psi_{-\vec{k}',\alpha=1_c}\psi_{\vec{k}',\beta=2_v}|V|\psi_{-\vec{k},\alpha=1_c}\psi_{\vec{k},\beta=2_v}} = 
\mbox{\ \ \ \ \ \ \ \ \ \ \ \ \ } \nonumber \\
F^{(H)}_{\vec{k},\alpha;\vec{k}',\beta}V^{(0)}(\bm{q}_{\parallel},q_z)
+F^{(0)}_{\vec{k},\alpha;\vec{k}',\beta}V^{(d)}(\bm{q}_{\parallel},q_z) ,
\end{eqnarray}
with $\vec{q}=\vec{k}-\vec{k}'$.  The form factors $F^{(H)}_{\vec{k},\alpha;\vec{k}',\beta}$
and $F^{(0)}_{\vec{k},\alpha;\vec{k}',\beta}$ are given in Eqs. (S4) of 
the Supplementary Material\cite{Supplemental3DTMD}. 

Equation \eqref{interintra} expresses the property that, 
due to the hybridization between the bands of the different type monolayers, 
there is a small intralayer contribution to the e-h potential.  This is because, while   
the electrons and holes in the hybridized bands are mostly in opposite layers, there is a small 
probability they will be in the same layer. 
At large momentum exchange $\bm{q}_{\parallel}$, the potential 
is dominated by 2D interactions between same type TMDs, $V^{(0)}(\bm{q}_{\parallel},q_z)$,
while at small $\bm{q}_{\parallel}$, 
the total interaction potential in Eq.\ \eqref{interintra} is dominated by 3D interactions 
between different type TMDs, $V^{(d)}(\bm{q}_{\parallel},q_z)$ (see Fig.\ S1 in the Supplementary Material\cite{Supplemental3DTMD}). 
Since pairing by the screened Coulomb attraction is primarily generated 
by two-particle scattering processes with small momentum exchange, pair formation is 3D in character.


Our interacting Hamiltonian for electrons and holes in the superlattice is, 
\begin{eqnarray}
\label{Grand-canonical-Hamiltonian}
&\mbox{\!\!\!\!}{\cal{H}}\! = \sum_{\vec{k} } ({\varepsilon}_{\vec{k},1_c}\!-\!\mu_e) c^{\dagger}_{\vec{k},1_c} c_{\vec{k},1_c}
\!\!\!+\!\! (-{\varepsilon}_{\vec{k},2_v}\!-\!\mu_h) d^{\dagger}_{\vec{k},2_v} d_{\vec{k},2_v}\!  \mbox{\!\!\!\!}
\nonumber \\
& + \displaystyle{\sum_{\substack{\vec{k}\vec{k}'}}} \!\!
 \braket{\psi_{-\vec{k}',1_c}\psi_{\vec{k}',2_v}\!|V|\psi_{-\vec{k},1_c}\psi_{\vec{k},2_v}}\mbox{\!}
c^{ \dagger}_{-\vec{k}',1_c}\mbox{\!}
d^{ \dagger}_{\vec{k}',2_v}\mbox{\!}
d_{\vec{k},2_v}\mbox{\!}
c_{-\vec{k},1_c} \nonumber \\
\end{eqnarray}
We make the standard transformation for the holes in the valence band 
to positively charged particles with positive energies, so the   
chemical potentials $\mu_e$ and $\mu_h$ in the monolayers are both positive.
$c^{\dagger}_{\vec{k},1_c}$ and $c_{\vec{k},1_c}$ ($d^{\dagger}_{\vec{k},2_v}$ and $d_{\vec{k},2_v}$) 
are the creation and destruction operators for the electrons (holes).  

The 3D superfluid gap $\Delta(\vec{k})$ at zero temperature is determined from the self-consistent mean-field equation,
\begin{equation}
\label{gapeq}
\Delta(\vec{k}) = -\sum_{\vec{k}'}V^{RPA}(\vec{k},\vec{k}')
\frac{\Delta(\vec{k}')}{2E_{\vec{k}'}} \ ,
\end{equation}
where $E_{\vec{k}} = \sqrt{\xi_{\vec{k}}^2+\Delta_{\vec{k}}^2}$, with  
$\xi_{\vec{k}} = \frac{1}{2}(\varepsilon_{\vec{k},1_c}-\varepsilon_{\vec{k},2_v})-\mu$.
We evaluate Eq.\ \eqref{gapeq} at a fixed value of the average chemical potential $\mu =\frac{1}{2} ({\mu_e+\mu_h})$.
The terms in the summation over $\bm{k}_{\parallel}'$ are  non-negligible only at low energies,
but the summation over $k_z'$ has significant contributions across the full  
Brillouin zone, {\it i.e.} between $\pm \pi/2d$.  
$V^{RPA}(\vec{k},\vec{k}')$  
is the self-consistent RPA screened e-h interaction in the superlattice 
in the presence of the superfluid.  The screening is due to 
the polarization of the electron and hole densities and the superfluid condensate\cite{Perali2013}. 
The expression for $V^{RPA}(\vec{k},\vec{k}')$  is given in the Supplementary Material\cite{Supplemental3DTMD}. 

For given values of the chemical potentials $\mu_e$ and $\mu_h$, 
the 3D electron and hole densities are given by,
\begin{equation}
\label{n_e}
n = \frac{g_v}{AN2d}\sum_{\vec{k}}\left(v_{\vec{k}}\right)^2 \ . 
\end{equation}
Note even though we set electron and hole densities $n$ equal, 
$\mu_e\neq \mu_h$ because of the unequal effective masses.

\begin{figure}
\includegraphics[trim=1.6cm 1cm 1.9cm 1.0cm, clip=true, angle=0,width=1\columnwidth]{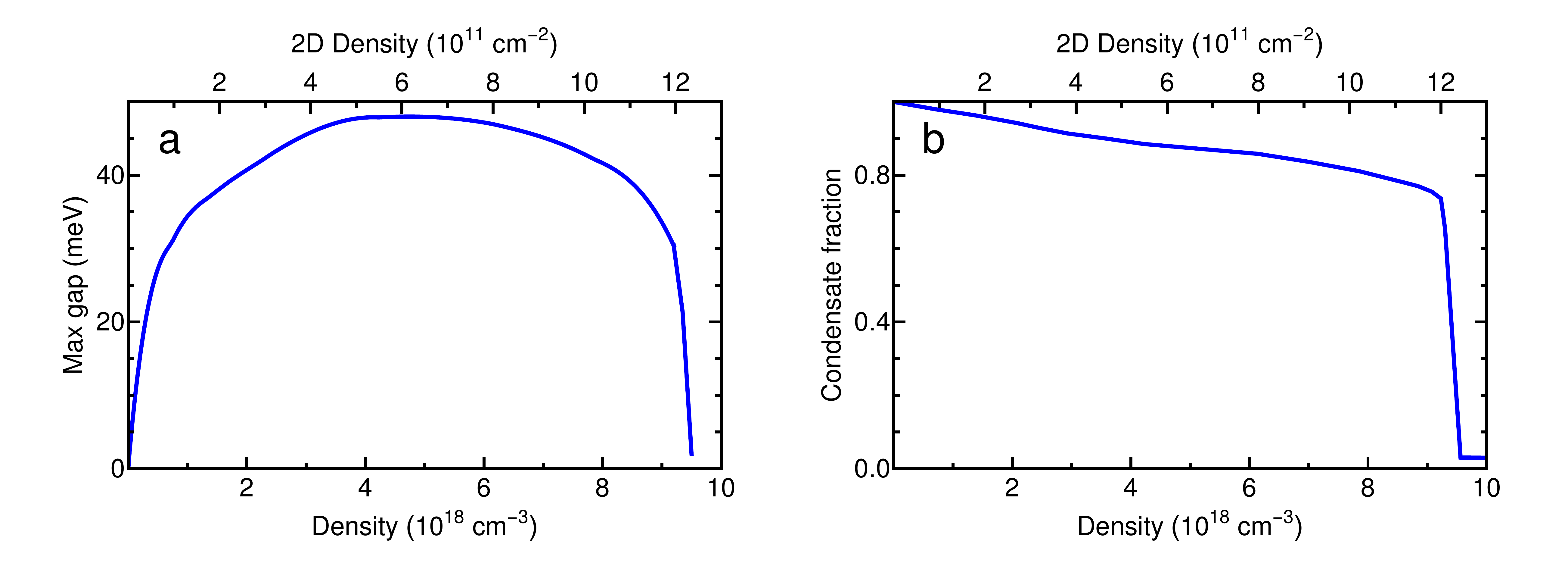}
\caption{(Color online) 
(a) Maximum superfluid gap ${\Delta^{max}}$ and  
(b) condensate fraction ${{\cal{C}}}$,
as functions of the equal electron and hole densities ${n}$.  
Top axis shows effective 2D density ${n_{2D}}$.
}
\label{Deltaandcandmu}
\end{figure}

Figure \ref{Deltaandcandmu}(a) shows the zero-temperature $\Delta^{max}$, the maximum of the momentum-dependent superfluid gap 
$\Delta(\vec{k})$ (Eq.\ \eqref{gapeq}), as a function of the 3D electron and hole densities $n$.  
For reference the top axis shows an effective 2D carrier density, defined as $n_{2D}=2dn$.  
At large densities, Coulomb screening suppresses the superfluidity. 
Below an onset density $n_0$, large gap superfluidity self-consistently weakens the screening sufficiently for 
superfluidity to appear.  As the density is further decreased, $\Delta^{max}$ 
increases to a maximum value of $48$ meV ($560$ K), and then decreases.  
Note that even for very small values of $n$, $\Delta^{max}$ remains in excess of $10$ meV ($120$ K).   
These large values of $\Delta^{max}$ reflect the strong e-h Coulomb pairing interaction.
Figure \ref{Deltaandcandmu}(b) shows the condensate fraction ${\cal{C}}$ 
that determines the density range for the BCS, BCS-BEC crossover, and BEC regimes
(see Eq.\ (S8) in the Supplementary Material\cite{Supplemental3DTMD}).  

%
\begin{figure}
\begin{center}\includegraphics[trim=1.6cm 1.0cm 0.1cm 1.0cm, clip=true, angle=0,width=0.8\columnwidth]{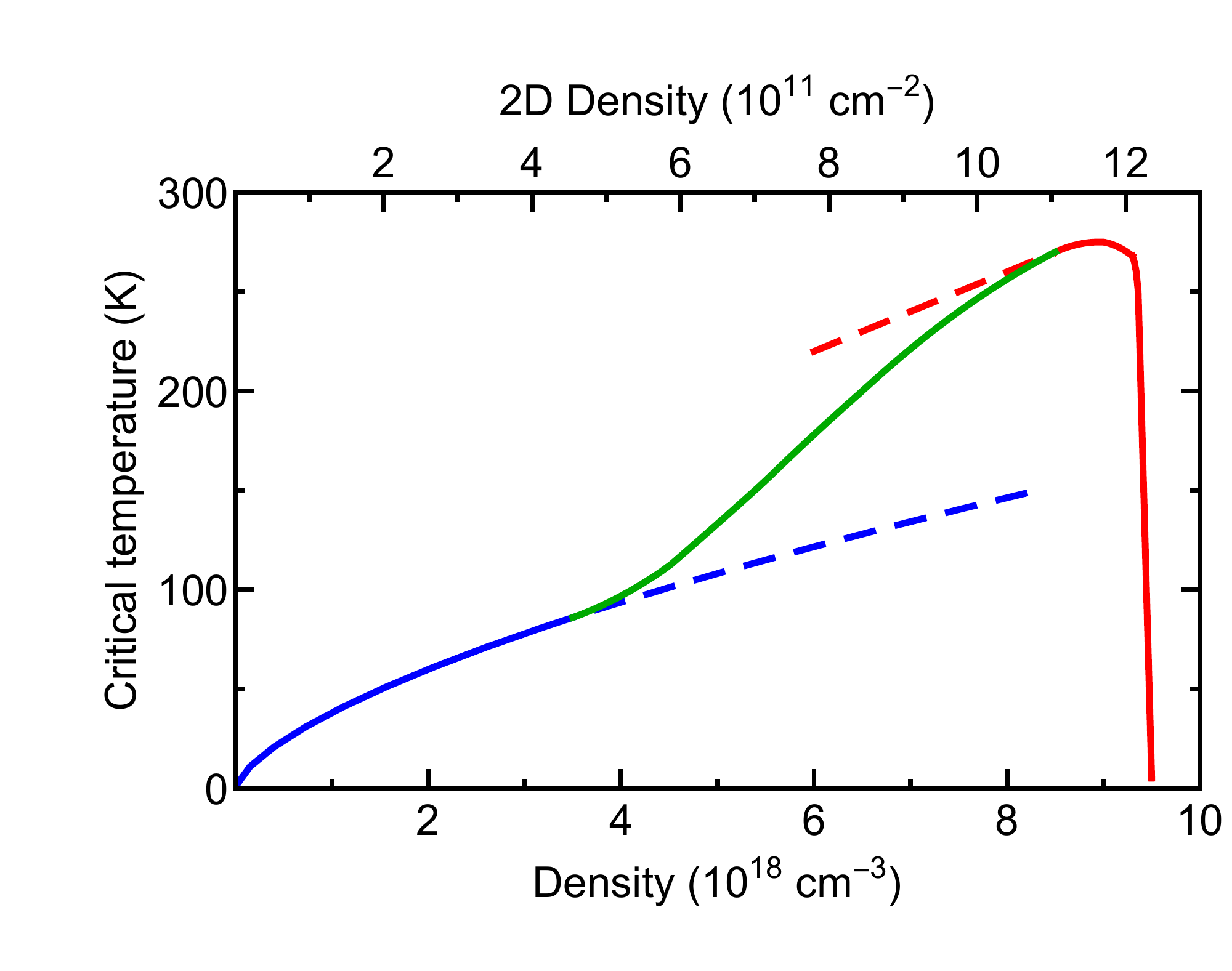}
\end{center}
\caption{(Color online) 
Superfluid transition temperature ${T_c}$ 
as a function of  ${n}$, the equal electron and hole density in the superlattice.     
Red line: ${T_c}$ determined in the BCS and BCS-BEC crossover regimes using Eqs.\ \eqref{gapeq} and \eqref{n_e} 
generalized to finite temperatures. 
Blue line: ${T_c}$ determined in the deep BEC regime using Eq.\ (S9) of the Supplementary Material\cite{Supplemental3DTMD}. 
Green line: interpolation.
}
\label{T_c}
\end{figure}

At high densities at weak-coupling, 
the superfluid transition temperature $T_c$ can be determined from the mean-field BCS equations,  
Eqs.\ \eqref{gapeq} and \eqref{n_e}, generalized to finite temperatures.  
As the density is lowered, we enter the BCS-BEC crossover regime.  With the increased pairing strength, 
the chemical potential $\mu$ must drop below the Fermi energy $E_F$ to keep the density fixed.  
This drop 
incorporates a large part of the effect of the fluctuations that build up as the crossover regime is penetrated.
Although within the crossover regime, the $T_c$ determined from the generalized 
Eqs.\ \eqref{gapeq} and \eqref{n_e} using the self-consistent $\mu$     
starts to overestimate the actual transition temperature, this overestimate 
is not expected to exceed $20$\% across the full crossover regime\cite{Haussmann1994,Pini2019}. 
For example, for ultra-cold fermions, the simplest non-selfconsistent
t-matrix approach overestimates the $T_c$ obtained by Quantum Monte Carlo (QMC) simulations
by only $\sim 20$\% at unitarity in the crossover regime (Fig.\ 3 of Ref.\ \citenum{Pini2019}).  
In this simplest t-matrix approach, the sole ingredient entering the $T_c$ calculation is 
the renormalization of the chemical potential.  
In the self-consistent screening, we retain the superfluid gap at zero $T$, 
since the pseudogap arising from the pair fluctuations should remain 
of the order of $\Delta(T\!\!=\!0)$ in the intermediate
coupling regime\cite{Perali2002}, and so to a large extent 
the low-lying excited states will continue to be excluded from the screening excitations,
suppressing the detrimental Coulomb screening.  
In this way we take into account a major part of the fluctuation effects that renormalize $T_c$ 
to lower values, by incorporating a large part of the fluctuations through the reduction of the chemical potential and 
through the development of the pseudogap. 

In the deep BEC regime at low densities (${\cal{C}} > 0.9$), this method for determining 
$T_c$ becomes unreliable, primarily because the pseudogap is replaced by a real gap 
of order of the pair binding energy.
In the deep BEC, we can approximate 
the e-h pairs as point-like bosons, so we can use the $T_c$ 
for BEC of non-interacting bosons (Eq.\ (S9) of the Supplementary Material\cite{Supplemental3DTMD}). 
The $T_c$ thus obtained is known to underestimate the actual $T_c$ for BEC 
as determined by QMC\cite{Burovski2008}.  
Finally, in the density range from the upper boundary of the BEC regime to  
the start of the deep BEC, we use a smooth interpolation of $T_c$ between the high- and low-density results.

Figure \ref{T_c} shows the resulting superfluid transition temperature in the superlattice.
In the deep-BEC regime, $T_c$ (blue curve) can approach $100$ K, many orders of magnitude
 larger than the BEC transition temperatures found in ultra-cold atom systems\cite{Anderson1995,Bradley1995,Davis1995}.  
These BEC transition temperatures are so much larger because the effective electron and hole masses are tiny compared to atomic masses, 
and because our densities are several orders of magnitudes larger than in ultra-cold atom systems.
Increasing the density causes $T_c$ to rapidly rise,  
pushing it to a maximum in the BCS-BEC crossover regime (red curve) 
very close to room temperature, $T_c=270$ K -- conveniently accessible in  a domestic refrigerator.  


While our calculations use the realistic band structure of a specific infinite superlattice, our conclusions remain
valid for finite superlattices consisting of more than a few monolayers.   
A further advantage of a 3D system over 2D systems is that it is much less susceptible to disorder, because 
percolation and screening favor 3D conduction.  Our results open the way to generating 
3D e-h superfluidity at room temperature in this and related superlattices.

\begin{acknowledgments} 
This work was supported by the Research Foundation of Flanders (FWO-Vl) 
through an aspirant research grant for MVDD, by the FLAG-ERA project TRANS-2D-TMD, 
and by the Australian Government through the Australian Research Council Centre of Excellence 
in Future Low-Energy Electronics (Project No. CE170100039).
We thank Milorad V. Milo\u{s}evi\'{c}, Pierbiagio Pieri and Jacques Tempere for helpful discussions.
\end{acknowledgments}


\begin{thebibliography}{35}
\bibitem [{Sup()}]{Supplemental3DTMD}%
  \BibitemOpen
  \href@noop {} {}\bibinfo {note} {{S}ee {S}upplementary {M}aterial, which
  includes {R}efs.\ [5],[18]}\BibitemShut {NoStop}%
\end{thebibliography}
\end{document}


\title{SUPPLEMENTARY MATERIAL \\
Three-Dimensional electron-hole superfluidity in a superlattice close to room temperature}

\author{M. Van der Donck}
\affiliation{Department of Physics, University of Antwerp, Groenenborgerlaan 171, 2020 Antwerp, Belgium}
\author{S. Conti}
\affiliation{Department of Physics, University of Antwerp, Groenenborgerlaan 171, 2020 Antwerp, Belgium}
\affiliation{Physics Division,  School of Science \& Technology, Universit\`a di Camerino, 62032 Camerino (MC), Italy}
\author{A. Perali}
\affiliation{Supernano Laboratory, School of Pharmacy, Universit\`a di Camerino, 62032 Camerino (MC), Italy}
\affiliation{ARC Centre of Excellence for Future Low Energy Electronics Technologies, 
School of Physics, The University of New South Wales, Sydney, N.S.W. 2052, Australia}
\author{A. R. Hamilton}
\affiliation{ARC Centre of Excellence for Future Low Energy Electronics Technologies, 
School of Physics, The University of New South Wales, Sydney, N.S.W. 2052, Australia}
\author{B. Partoens}
\affiliation{Department of Physics, University of Antwerp, Groenenborgerlaan 171, 2020 Antwerp, Belgium}
\author{F. M. Peeters}
\affiliation{Department of Physics, University of Antwerp, Groenenborgerlaan 171, 2020 Antwerp, Belgium}
\author{D. Neilson}
\affiliation{Department of Physics, University of Antwerp, Groenenborgerlaan 171, 2020 Antwerp, Belgium}
\affiliation{ARC Centre of Excellence for Future Low Energy Electronics Technologies, 
School of Physics, The University of New South Wales, Sydney, N.S.W. 2052, Australia}

\begin{abstract}
In this supplementary material we provide a table with relevant parameters for the two TMDs in our quantitative example. We explain how the bare electron-hole interaction potential can be derived, taking into account the effects of the superlattice geometry and of hybridization between the TMDs. Next, we show how screening effects in the presence of the superfluid are included by means of the RPA. Finally, expressions for the condensate fraction and the critical temperature for non-interacting bosons are given.
\end{abstract}
 
\maketitle
\newcommand{\beginsupplement}{%
        \setcounter{table}{0}
        \renewcommand{\thetable}{S\arabic{table}}%
        \setcounter{figure}{0}
        \renewcommand{\thefigure}{S\arabic{figure}}%
        \setcounter{section}{0}
        \renewcommand{\thesection}{S\arabic{section}}%
        \setcounter{equation}{0}
        \renewcommand{\theequation}{S\arabic{equation}}%
  }
 \beginsupplement
      \renewcommand{\citenumfont}[1]{S#1}

\bibpunct{[}{]}{;}{n}{}{}

\section{Table of Parameters for WS$_{2}$ and WSe$_{2}$}

\begin{table}[h]
\centering
\begin{tabular}{|c c c c c c c|}
\hline
\hline
 & $a$ (nm) & $t$ (eV) & $E_g$ (eV) & $2\lambda_c$ (eV) & $2\lambda_v$ (eV) & $2t_v$ (eV) \\
\hline
\hline
W$\text{S}_2$ & 0.32 & 1.37 & 1.79 & 0.027 & 0.43 & 0.109 \\
\hline
WS$\text{e}_2$ & 0.33 & 1.19 & 1.60 & 0.038 & 0.46 & 0.134 \\
\hline
\hline
\end{tabular}
\caption{\textbf{
Parameters for WS$_{\bm 2}$ and WSe$_{\bm 2}$:
lattice constant\cite{Xiao2012} (${\bm a}$), hopping parameter\cite{Xiao2012} (${\bm t}$), 
band gap\cite{Xiao2012} ($\bm{E_g}$), spin splitting of conduction band\cite{Kosmider2013}  ($\bm{2\lambda_c}$) 
and valence band\cite{Zhu2011}  ($\bm{2\lambda_v}$), interlayer hopping parameter\cite{Gong2013} 
($\bm{t_v}$).
}
}
\label{table:mattable}
\end{table}

\section{E-h interactions: Effects of superlattice geometry and hybridization}

The energy spectrum (Fig.\ 1 in the manuscript) and eigenstates are obtained by numerically solving 
the eigenvalue equation of the $4\times 4$ Hamiltonian (Eq.\ (1) in the manuscript). 
For given spin and valley quantum numbers, the single-particle eigenstate for energy band 
$\beta$ is $\ket{\psi_{\vec{k},\beta}}$, which 
can be written as the four-component vector, 
\begin{equation}
\label{1pspinor}
\ket{\psi_{\vec{k},\beta}} = \begin{pmatrix}
C_{1_c,\beta}^{\vec{k}}\ket{\Phi_{\vec{k},\ell=1}} \\ 
C_{1_v,\beta}^{\vec{k}}\ket{\Phi_{\vec{k},\ell=1}} \\
C_{2_c,\beta}^{\vec{k}}\ket{\Phi_{\vec{k},\ell=2}} \\
C_{2_v,\beta}^{\vec{k}}\ket{\Phi_{\vec{k},\ell=2}} 
\end{pmatrix} \ .
\end{equation}
The different pseudospin states $\ket{\Phi_{\vec{k},\ell}}$ are defined below. The weighting coefficients $C_{i,\beta}^{\vec{k}}$  
include both the effects of interlayer hopping generated by $ {\cal{T}}(k_z)$, 
and the hybridization of the conduction and valence bands.
We may assume the continuum approximation for the dispersion of the bands 
in the parallel direction at low energies, but in the $z$-direction, with its small band widths,  
all $k_z$-values in the first Brillouin zone must be considered. 
\begin{figure}[h]
\begin{center}
\includegraphics[angle=0,width=0.9\columnwidth]{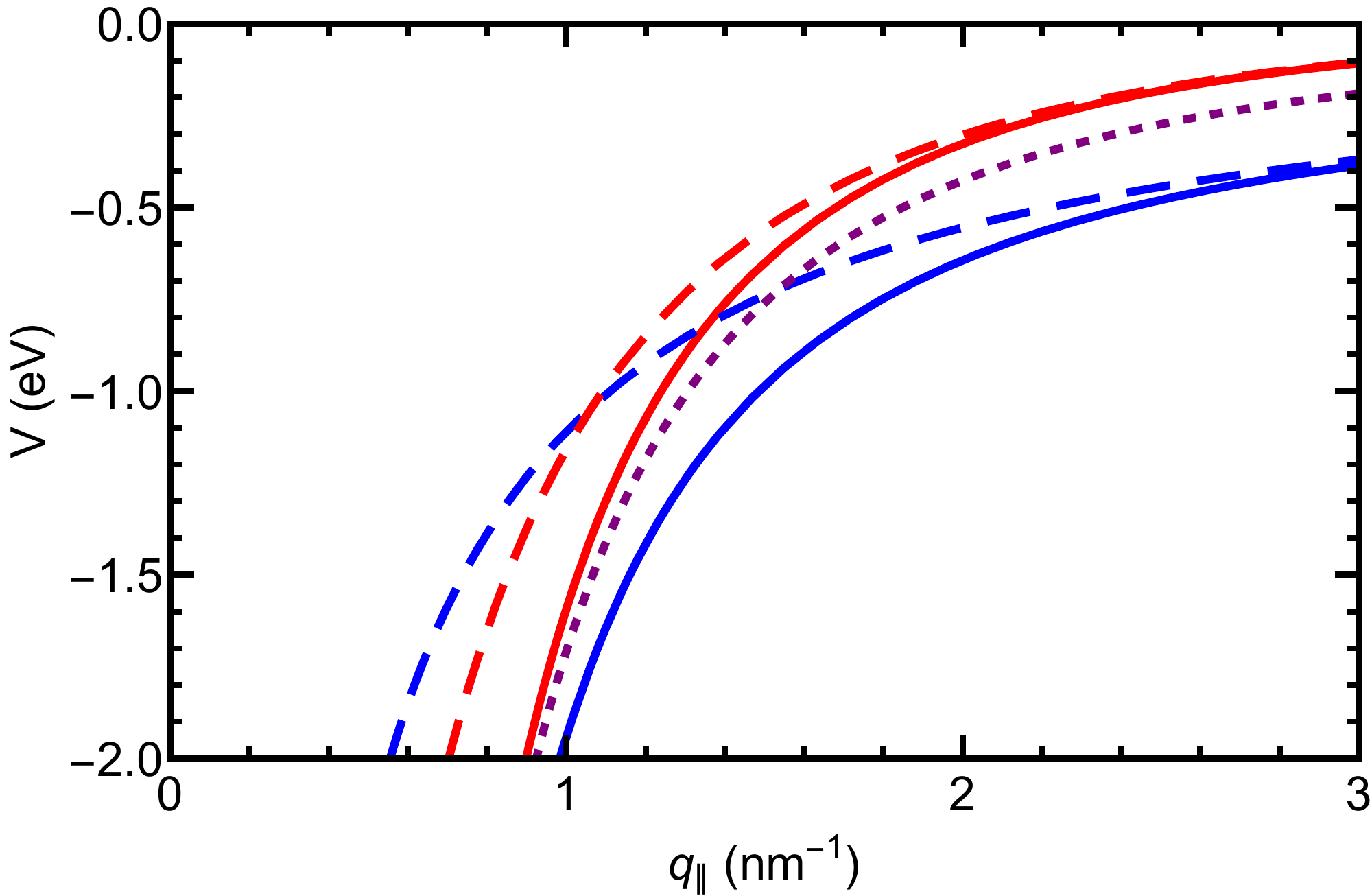}
\end{center}
\caption{(Color online)
Solid blue curve: Intralayer interaction potential $V^{(0)}(\bm{q}_{\parallel},q_z\!\!=\!0)$   (Eq.\ (5) in the main manuscript)
as a function of $\bm{q_{\parallel}}$.  
Solid red curve: interlayer interaction potential $V^{(d)}(\bm{q}_{\parallel},q_z\!\!=\!0)$  (Eq.\ (6) in the main manuscript).
Dashed blue curve: 2D intralayer  interaction potential (${\propto 1/q_{\parallel}}$).
Dashed red curve:  2D interlayer  interaction potential (${\propto 2e^{-q_{\parallel}d}/q_{\parallel}}$).
Dotted purple curve: 3D interaction potential (${\propto 1/q_{\parallel}^2}$).
}
\label{Potential}
\end{figure}

Since the influence of the interlayer hopping on the energy bands is small because of the energy mismatch 
between the bands of the different TMDs, we can write, 
\begin{equation}
\label{wavefunction}
\braket{\vec{r}\,|\Phi_{\vec{k},\ell}} \!=\! \frac{1}{\sqrt{NA}}e^{i\mathbf{k}_{\parallel}\cdot\bm{r}_{\parallel}}
\!\!\!\!\!\!\!\
\sum_{j=-N/2}^{N/2} \!\!\!\! \delta^{1/2}(z-j 2d-z_\ell d)e^{i(2j+z_\ell)k_zd} \ ,
\end{equation}
with $N$ the number of TMD heterostructures and $z_{\ell=1}=0$ ($z_{\ell=2}=1$) representing the relative position in the $z$-direction of each 
WS$_2$ (WSe$_2$) monolayer in the superlattice.  

The matrix element of the interaction potential between these states is given by
\begin{eqnarray}
\label{Vq}
&\!\!\!\!\!\!\!\!\!\!\!\!\!\!\!\!\!\!\!\!\!\!\!\!\!\!\!\!\!\!\!\!\!\!\!\!\!\!\!\!\!\!\!\!\!\!\!\!\!\!\!\!\!\!\!\!\!
\braket{\Phi_{\vec{\kappa}',\ell_2'}\Phi_{\vec{k}',\ell_1'}|V|\Phi_{\vec{\kappa},\ell_2}\Phi_{\vec{k},\ell_1}} = 
 \nonumber \\
&-\delta_{\ell_1,l_1'} \delta_{\ell_2,l_2'} \delta_{\vec{k}+\vec{\kappa},\vec{k}'+\vec{\kappa}'}
\left[ \frac{e^2}{4\pi\varepsilon_r\varepsilon_0NA}
\frac{2\pi}{q_{\parallel}} \right] \times
 \nonumber \\
&\ \ \ \ \ \ \ \ \ \ \displaystyle{\sum_{w=-N}^{N}} e^{i(2w-z_{\ell_1}+z_{\ell_2})q_zd}
e^{-|2w-z_{\ell_1}+z_{\ell_2}|q_{\parallel}d} \ ,
\end{eqnarray}
with $\vec{q}=\vec{k}-\vec{k}'=\vec{\kappa}'-\vec{\kappa}$.  
The factor $\delta_{\ell_1,\ell_1'}\delta_{\ell_2,\ell_2'} $ 
confines the electrons and holes to their original monolayers when they scatter.  
For $N\rightarrow\infty$ the summation leads to $V^{(0)}$, Eq.\ (4) in the manuscript for $\ell_1=\ell_2$, i.e. between same type TMD monolayers, and to $V^{(d)}$, Eq.\ (5) for $\ell_1\neq\ell_2$, i.e. between different type TMD monolayers. These interaction potentials are shown in Fig. \ref{Potential}.

Evaluating the interaction potential between the eigenstates of Eq. \eqref{1pspinor} leads to Eq.\ (6) in the manuscript. The form factors appearing in Eq.\ (6) are given by,
\begin{equation}
\label{form}
\begin{split}
&F^{(0)}_{\vec{k},\alpha;\vec{k}',\beta} = 
\mathcal{C}^{1,\alpha}_{\vec{k},\vec{k}'}\mathcal{C}^{2,\beta}_{\vec{k},\vec{k}'}
+\mathcal{C}^{2,\alpha}_{\vec{k},\vec{k}'}\mathcal{C}^{1,\beta}_{\vec{k},\vec{k}'}\ , \\
&F^{(H)}_{\vec{k},\alpha;\vec{k}',\beta} = 
\mathcal{C}^{1,\alpha}_{\vec{k},\vec{k}'}\mathcal{C}^{1,\beta}_{\vec{k},\vec{k}'}
+\mathcal{C}^{2,\alpha}_{\vec{k},\vec{k}'}\mathcal{C}^{2,\beta}_{\vec{k},\vec{k}'}\ , 
\end{split}
\end{equation}
with $\mathcal{C}^{\ell,\alpha}_{\vec{k},\vec{k}'} \equiv 
\sum_{j=c,v} (C_{\ell_j,\alpha}^{\vec{k}'})^\star C_{\ell_j,\alpha}^{\vec{k}}$.
From Eq.\ \eqref{form}, we can see for $\alpha=1_c$ and $\beta=2_v$, that 
$F^{(0)}_{\vec{k},\alpha;\vec{k}',\beta}$ will be large,
and that the hybridized $F^{(H)}_{\vec{k},\alpha;\vec{k}',\beta} $ will be small.

\begin{widetext}
\section{RPA screening in the superlattice in the presence of the superfluid}

$V^{RPA}(\vec{k},\vec{k}')$, appearing in Eq.\ (8) in the manuscript,
is the self-consistent RPA screened e-h interaction in the superlattice 
in the presence of the superfluid.  The screening is due to 
the polarization of the electron and hole densities and the superfluid condensate\cite{Perali2013}.  
It is given by,
\begin{equation}
\label{RPAinter}
V^{RPA}(\vec{k},\vec{k}') =
\frac{F^{(0)}_{\vec{k},\alpha=1_c;\vec{k}',\beta=2_v}V^{(d)}(\vec{q})
+F^{(H)}_{\vec{k},\alpha=1_c;\vec{k}',\beta=2_v}V^{(0)}(\vec{q})}
{1+2V^{(0)}(\vec{q}) \left[\Pi^{(0)}_n(\vec{q})+\Pi^{(H)}_a(\vec{q})\right]
+2V^{(d)}(\vec{q}) \left[\Pi^{(H)}_n(\vec{q})+\Pi^{(0)}_a(\vec{q})\right]},
\end{equation}
with $\vec{q}=\vec{k}-\vec{k}'$.  The presence of the superfluid strongly affects 
the $\Pi$ polarization functions\cite{Perali2013} in Eq.\ \eqref{RPAinter},
which for the superlattice are defined as,
\begin{eqnarray}
\label{pi}
\Pi^{(\lambda)}_n(\vec{q}) &=& -g_v\sum_{\vec{k}}
\frac{F^{(\lambda)}_{\vec{k}+\vec{q},\alpha=1_c;\vec{k},\beta=2_v}}{E_{\vec{k}+\vec{q}} + E_{\vec{k}}}  
\left\{\left(u_{\vec{k}+\vec{q}}v_{\vec{k}}\right)^2+\left(u_{\vec{k}}v_{\vec{k}+\vec{q}}\right)^2\right\} \\
\Pi^{(\lambda)}_a(\vec{q}) &=&  g_v\sum_{\vec{k}}
\frac{F^{(\lambda)}_{\vec{k}+\vec{q},\alpha=1_c;\vec{k},\beta=2_v}}
{E_{\vec{k}+\vec{q}}+E_{\vec{k}}}\left\{2u_{\vec{k}+\vec{q}}v_{\vec{k}}u_{\vec{k}}v_{\vec{k}+\vec{q}}\right\}, 
\end{eqnarray}
where $ (\lambda)=(0),(H)$ (recall Eq.\ \eqref{form}).  
The Bogoliubov amplitudes are $u_{\vec{k}}^2 = \tfrac{1}{2}\left(1+{\xi_{\vec{k}}}/{E_{\vec{k}}}\right)$ and  
$v_{\vec{k}}^2 = \tfrac{1}{2}\left(1-{\xi_{\vec{k}}}/{E_{\vec{k}}}\right)$.
\end{widetext}

\section{Condensate fraction}

The condensate fraction,
\begin{equation}
\label{cfraction}
{\cal{C}} = \frac{\sum_{\vec{k}}(u_{\vec{k}}v_{\vec{k}})^2}{\sum_{\vec{k}}(v_{\vec{k}})^2} \ ,
\end{equation}
measures the fraction of carriers in the condensate\cite{Yang1962,Regal2004,Perali2005,Manini2005}.  
${\cal{C}}$ characterizes the different regimes of pairing in ultra-cold  fermions\cite{Salasnich2005},
and we apply the same criterion:
in the BCS regime ${\cal{C}} < 0.2$, with only a small fraction of the electrons and holes close to the Fermi surface 
forming  pairs and condensing;
$0.2< {\cal{C}} < 0.8$ characterizes the BCS-BEC crossover regime; 
in the BEC regime ${\cal{C}} > 0.8$, and most carriers have formed bosonic pairs and condensed; 
in the deep BEC regime, ${\cal{C}} > 0.9$, the condensed bosonic pairs are compact and very weakly interacting, and there are almost no free carriers.  

\section{$T_c$ for BEC of non-interacting bosons}

The $T_c$ for Bose-Einstein Condensation of non-interacting bosons is determined by inverting the equation\cite{Pethick2008}, 
\begin{equation}
\label{TcBEC}
n = \frac{2}{AN2d}\sum_{\vec{k}}\frac{1}{e^{\left(\varepsilon_{\vec{k},1_c}-\varepsilon_{\vec{k},2_v}-\varepsilon_{\vec{0},1_c}
+\varepsilon_{\vec{0},2_v}\right)/(k_B T_c)}-1} \ .
\end{equation}

\makeatletter
\renewcommand\@biblabel[1]{[S#1]}